\documentclass[twocolumn,twocolumn]{IEEEtran}
\usepackage[T1]{fontenc}
\usepackage[latin9]{inputenc}
\usepackage{color}
\usepackage{array}
\usepackage{float}
\usepackage{amsmath}
\usepackage{amsthm}
\usepackage{graphicx}
\usepackage[unicode=true,
 bookmarks=true,bookmarksnumbered=true,bookmarksopen=true,bookmarksopenlevel=1,
 breaklinks=false,pdfborder={0 0 0},pdfborderstyle={},backref=false,colorlinks=false]
 {hyperref}
\hypersetup{pdftitle={Your Title},
 pdfauthor={Your Name},
 pdfpagelayout=OneColumn, pdfnewwindow=true, pdfstartview=XYZ, plainpages=false}
\usepackage{breakurl}

\makeatletter

\providecommand{\tabularnewline}{\\}
\floatstyle{ruled}
\newfloat{algorithm}{tbp}{loa}
\providecommand{\algorithmname}{Algorithm}
\floatname{algorithm}{\protect\algorithmname}

  \theoremstyle{definition}
  \newtheorem{defn}{\protect\definitionname}

\usepackage[caption=false,font=footnotesize]{subfig}
\usepackage{algorithm}
\usepackage{algorithmic}

\usepackage{multirow} 
\usepackage{amsmath} 
\usepackage{xcolor}

\allowdisplaybreaks[4]

\ifCLASSOPTIONcompsoc
\usepackage[caption=false,font=normalsize,labelfont=sf,textfont=sf]{subfig}
\else
\usepackage[caption=false,font=footnotesize]{subfig}
\fi

\usepackage{cite}
\usepackage{bm}
\usepackage{algorithmic}
\usepackage{algorithm}
\usepackage{graphicx}
\IEEEoverridecommandlockouts

\usepackage{lettrine}

\@ifundefined{showcaptionsetup}{}{%
 \PassOptionsToPackage{caption=false}{subfig}}
\usepackage{subfig}
\makeatother

\providecommand{\definitionname}{Definition}

\begin{document}

\title{\textcolor{black}{Hierarchical Aerial Computing for Internet of Things
via Cooperation of HAPs and UAVs}}

\author{Ziye Jia, \IEEEmembership{Member,~IEEE}, Qihui Wu, \IEEEmembership{Senior Member,~IEEE},
Chao Dong, \IEEEmembership{Member,~IEEE},\\ Chau Yuen, \IEEEmembership{Fellow,~IEEE},
and Zhu Han, \IEEEmembership{Fellow,~IEEE},\thanks{Ziye Jia, Qhui Wu and Chao Dong are with the College of Electronic
and Information Engineering, Nanjing University of Aeronautics and
Astronautics, Nanjing 210000, China, (e-mail: jiaziye@nuaa.edu.cn,
wuqihui@nuaa.edu.cn, dch@nuaa.edu.cn).

Chau Yuen is with the Engineering Product Development Pillar, Singapore
University of Technology and Design, Singapore (e-mail: yuenchau@sutd.edu.sg).

Zhu Han is with the University of Houston, TX 77004, USA (e-mail:
zhan2@uh.edu), and also with the Department of Computer Science and
Engineering, Kyung Hee University, Seoul, 446-701, South Korea.}}
\maketitle
\begin{abstract}
\textcolor{black}{With the explosive increment of computation requirements,
the multi-access edge computing (MEC) paradigm appears as an effective
mechanism. Besides, as for the Internet of Things (IoT) in disasters
or remote areas requiring MEC services, unmanned aerial vehicles (UAVs)
and high altitude platforms (HAPs) are available to provide aerial
computing services for these IoT devices. In this paper, we develop
the hierarchical aerial computing framework composed of HAPs and UAVs,
to provide MEC services for various IoT applications. In particular,
the problem is formulated to maximize the total IoT data computed
by the aerial MEC platforms, restricted by the delay requirement of
IoT and multiple resource constraints of UAVs and HAPs, which is an
integer programming problem and intractable to solve. Due to the }prohibitive\textcolor{black}{{}
complexity of exhaustive search, we handle the problem by presenting
the matching game theory based algorithm to deal with the offloading
decisions from IoT devices to UAVs, as well as a heuristic algorithm
for the offloading decisions between UAVs and HAPs. The external effect
affected by interplay of different IoT devices in the matching is
tackled by the externality elimination mechanism. Besides, an adjustment
algorithm is also proposed to make the best of aerial resources. The
complexity of proposed algorithms is analyzed and extensive simulation
results verify the efficiency of the proposed algorithms, and the
system performances are also analyzed by the numerical results.}
\end{abstract}


\section{Introduction}

\lettrine[lines=2]{A}{S} the advent and development of\textcolor{black}{{}
the sixth-generation wireless systems (6G), the issue related to}\textbf{\textcolor{black}{{}
}}\textcolor{black}{Internet of Things (IoT) has attracted more and
more attentions, due to the explosive increment of IoT devices, such
as surveillance camera, smart wearable devices, smart framing, and
the IoT equipments in disasters or remote areas }\cite{529-6gMag,967-6G-survey-Niyato}\textcolor{black}{.
}Most IoT applications have requirements of intensive computation
with delay restriction. However, IoT devices are typically equipped
with limited computing and energy resources, which restrict the intensive
computation demand being completed locally by IoT \cite{1060-multilayer-MEC-IoT}.
Fortunately, the advent of\textcolor{black}{{} multi-access edge computing
(MEC) paradigm provides an effective} mechanism to help IoT tackle
the computation tasks \cite{1032-6g-aerialAccess,1085-MEC-sky,1055-MEC-MIMO-VTMAG,1109-V2X-offloading}.
Since the IoT devices in remote areas or emergency circumstances lack
services from terrestrial cellular networks, the platforms in the
aerial access network (AAN) such as\textcolor{black}{{} high altitude
platforms (HAPs) and unmanned aerial vehicles (UAVs)} equipped with
\textcolor{black}{computation} resources\textcolor{black}{{} }are introduced
as\textcolor{black}{{} effective MEC }candidates \cite{884-HAP-survey2021,1079-HAP-3C-MAG,1110-UAV-zhaonan,1111-UAV-adHoc}.
Although both HAPs and UAVs in AAN can extend the connectivity for
IoT devices, they are characterized by the different flight height,
load capacity, and endurance time. The cooperation of HAPs and UAVs
can provide powerful MEC services for terrestrial IoT devices \cite{ziyejia-jsac,1083-UAV-3D-jsac2021,1059-UAV-FL-Comm,751-UAV-zhaonan}.

Generally, HAPs can endure at a fixed position around the altitude
of 20km for several months, which can serve as stable base stations
in the air due to large coverages and powerful payloads \cite{990-HAP-mag2021,jzy-globecom2021}.
Accordingly, HAPs can provide large and stable coverage for both terrestrial
IoT devices and UAVs in low altitude. Besides, HAPs can carry powerful
loading equipments such as computing devices and batteries. There
exist significant researches for HAPs in industry. For example, the
solar HAPs developed by HAPSMobile aim to provide network services
in the sky \cite{HAPS-MOBILE,jzy_IoTJ_bender}. However, the direct
connection to HAPs by IoT devices with limited power supply is unacceptable
for the delay requirement. Alternatively, compared with HAPs, the
advantage of UAVs is addressed by flexible flight with low altitude,
and the rotary-wing UAV is able to float at a quasi-static position
for a couple of hours. Consequently, UAVs can provide available access
for the ground IoT devices due to the possible proximity \cite{1020-dongchao-mag-uav,jzy-IoTJ2020,743-drone-mag}.
However, UAVs' resources (e.g. computation, energy, and transmission
power) and endurance time are limited due to the small carrying capacity,
and the IoT data offloaded on the UAV may not be satisfied within
the tolerant delay \cite{BOOK-UAV-hongliang}. In this account, the
cooperation of HAPs and UAVs to provide MEC services for IoT is necessary,
in which UAVs play two roles: completing the lightweight computation
IoT tasks, and relay other IoT data to HAPs for MEC services.\textcolor{blue}{{} }

\begin{figure}[t]
\centering

\includegraphics[scale=0.42]{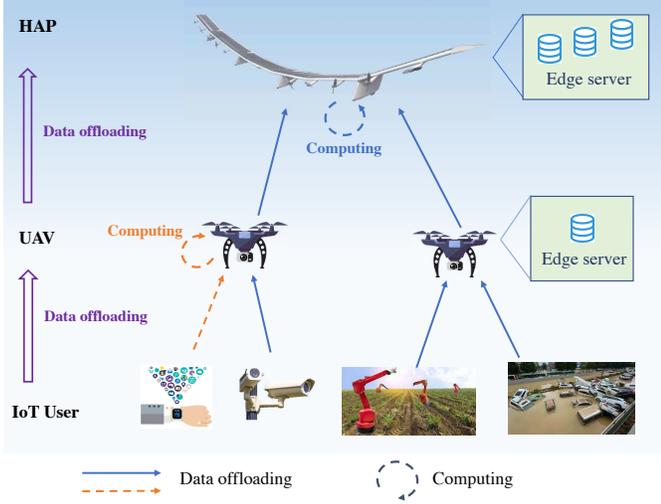}

\caption{\textcolor{black}{Hierarchical} aerial computing framework.\label{fig:scenario}}
\end{figure}

\textcolor{black}{In this paper, we propose the hierarchical aerial
computing framework, as shown in Fig. \ref{fig:scenario}, which is
composed of HAPs and UAVs in the air to provide MEC services for the
terrestrial IoT devices. }Specifically, an IoT device can offload
the computation demands to a UAV, and the data can be processed by
the UAV if the total time cost, including transmission and computation,
can meet the IoT's delay requirement. Otherwise, as for the heavy
computing IoT demands, the IoT data can be relayed by a UAV to a HAP
and leverage the HAP's powerful computation capacity. Notice here,
the offloading decisions have a trade-off between UAVs and HAPs if
the aerial resources are abundant. Taking all of these issues into
account, we focus on maximizing the total data being successfully
computed by UAVs and HAPs, constrained by multiple resources limitations
as well as integer decision restrictions. The problem is in the form
of integer programming, and is intractable to obtain an effective
solution, due to the prohibitive\textcolor{black}{{} complexity} of
the exhaustive search, especially in a large scale network.

To address the challenge for solutions based on the above discussion
and inspired by the matching game theory \cite{BOOK-matching}, we
primarily adopt the matching game based algorithm to deal with the
data offloading decision from IoT devices to UAVs. Therein, the preference
list construction for the participants is the key issue, since the
objective as well as a couple of constraints of the original problem
need to be implied in the preference lists \cite{BOOK-Roth}. In addition,
since the offloading decisions from different IoT device\textcolor{black}{s
may give }rise to the variation of IoT's preference lists, which is
termed as the external effect. In order to handle the issue of preference
list variation, we further present the externality elimination algorithm
to re-stabilize the matching between IoT devices and UAVs. In terms
of the data offloading decision from UAVs to HAPs, we propose a heuristic
algorithm to satisfy more IoT with rigorous delay restriction. \textcolor{black}{Moreover,
after the data offloading from UAVs to HAPs, UAVs may have redundant
resources. In this case, if there still exist unserved IoT devices,
we further design the adjustment algorithm to take full advantage
of aerial resources.}

Taking all the above discussions into account, the main contributions
of this paper are summarized as below. 
\begin{itemize}
\item We propose the hierarchical aerial computing framework composed of
HAPs and UAVs. Both HAPs and UAVs can provide the MEC service for
the terrestrial IoT devices, while HAPs have powerful computing and
energy payloads, which assist UAVs to complete the computing intensive
tasks. Besides, the detailed problem is formulated to maximize the
total successful computed data, constrained by multiple resource limitations,
and binary contact restriction.
\item Due to the\textcolor{black}{{} }prohibitive\textcolor{black}{{} complexity
to directly solve the formulated problem, we tackle the problem into
two stages. We present the matching game based algorithm as well as
the externality elimination algorithm to handle the data offloading
problem from IoTs to UAVs in the first stage, and a heuristic algorithm
for the data offloading problem from UAVs to HAPs. Besides, an adjustment
algorithm is further proposed to optimize the usage of aerial resources.
The time complexity of the proposed algorithms is also analyzed.}
\item Simulations are conducted and verify the efficiency of proposed algorithms,
and the effect of different algorithms are also evaluated from the
numerical results. Besides, the influence of system parameters such
as the computation ability of HAPs and UAVs are also analyzed.
\end{itemize}
\par The remainder of the paper is arranged as follows. In Section
\ref{sec:Literature-Review}, the literature review of recent related
works is discussed. We present the system model and corresponding
problem formulation in Section \ref{sec:System-Model}. The specific
algorithm design is proposed in Section \ref{sec:Algorithm-Design},
followed by the numerical results and performance evaluation in Section
\ref{sec:Performance-Evaluation}. Finally, the paper is concluded
in Section \ref{sec:Conclusions}.

\section{Literature Review\label{sec:Literature-Review}}

As for the UAV based aerial computing, there exist abundant related
works. For example, in \cite{1051-UAV-MEC-zhoufuhui}, the authors
have investigated the coupling of MEC and wireless power transfer
on UAVs, and two computation offloading modes including the partial
and binary modes have been considered. The problem has focused on
maximizing the total weighted computation rate by optimizing multiple
metrics such as transmission power, offloading times and trajectory
of the UAV. \cite{1052-UAV-MEC-energy} has jointly optimized the
total energy consumption of UAVs and users in the multiple UAV-enabled
MEC networks, considering the latency requirement and UAV location
planning. In \cite{1054-UAV-MEC-IoT}, the authors have provided the
multi-UAV enabled MEC framework for IoT computation offloading, and
both the processing efficiency and load balance have been considered
to optimize the network design. The authors in \cite{1058-UAV-MEC-yangbo}
have focused on the edge computing on UAVs to identify a mobile target
and keep tracking, considering the stringent and accurate latency
requirement, and a tradeoff has been obtained between the total cost
and inference error. \cite{1064-UAV-MEC-IoT} has proposed the UAV
assisted MEC framework for the time-sensitive IoT users, and the number
of successful served IoT devices and the resource-efficient UAV trajectory
has been coupled to been optimized. A reconnaissance task selection
and scheduling by the UAV-based MEC structure has been investigated
in \cite{1088-UAV-MEC-timevaring}, in which the reconnaissance task
has time-varying priority, and the total reconnaissance utility has
been maximized in the optimization problem. In \cite{1062-UAV-MEC-game-wuqihui},
the authors have investigated the computation offloading optimization
of UAVs in different layers by combining the channel allocation and
position scheduling as well, in which the Stackelberg game has been
employed to model the leader and follower relations between the two-layer
UAVs.

\textcolor{black}{Different with }UAVs, HAPs are characterized by
higher flight altitude and stronger payload, so that HAPs can provide
intensive computing services. \textcolor{black}{A couple of recent
works with respect to }the HAP-based aerial computing have been presented.
For example, \cite{1047-HAP-FL} has focused on the task computation
in the computing-enabled high-altitude balloons, which are deemed
as wireless base stations, and the federated learning based algorithm
has been designed to minimize the energy and time consumption during
the data offloading procedure. In \cite{1081-HAPnet-MEC}, a network
composed HAPs to provide massive access and edge computing services
has been presented, aiming to guarantee efficient connection and low
latency for massive IoT users. \cite{1091-caching-MEC-HAP} has proposed
a HAP based caching and computation offloading framework to improve
the latency of intelligent transportation systems, and a reinforcement
learning mechanism has been designed to tackle the corresponding mixed
integer nonlinear programming problem with efficiency. The authors
in \cite{1050-yangyang-MEC-HAP} have presented the computation offloading
structure in the HAPs-MEC-cloud networks, as the computing, communication
and caching resource allocation problem with intractability, and a
column generation based algorithm has been designed to handle the
problem. 

As for the multiple layers of computation platforms in the air, \cite{984-HAP-MEC}
has proposed a MEC architecture composed of drones and HAPs, providing
both radio access and computing tasks for the terrestrial users, and
the concept of end-to-end slice has also been presented as well as
the logic architecture of the user-drone-HAP system. In \cite{979-HAP-zhanglong},
the authors have focused on the data offloading in the space-air-ground
networks, in which HAPs serve the aerial computing platforms to complete
the MEC tasks, and the corresponding problem is formulated to maximize
the sum data rate, and is tackled by the hypergraph based mechanism.
\cite{1067-SAG-MEC} has proposed a space-air-ground enabled edge-cloud
computing framework composed of UAVs and satellites, in which UAVs
can serve the low-delay MEC requirement while satellites enable ubiquitous
cloud computing. The authors in \cite{1066-SAG-MEC-learning} have
presented the space-air-ground networks with MEC and cloud computing
for data offloading, and the Lyapunov based mechanism has been employed
to tackle the queue-aware optimization problem.

\textcolor{black}{With the above discussions with respect to aerial
computing, there exists a couple of works related to UAVs and HAPs.
However, to the extent of our knowledge, as for the cooperation of
UAVs and HAPs to provide the hierarchical MEC service for IoT, the
detailed cooperation model as well as corresponding schemes have not
been investigated. Hence, in this work, the issue of how to efficiently
leverage the hierarchical aerial resources of UAVs and HAPs will be
addressed. }

\section{System Model and Problem Formulation\label{sec:System-Model}}

In this section, we firstly present the system model in detail, including
the hierarchical aerial computing scenario in Section \ref{subsec:Hierarchical-scenario},
the communication model in Section \ref{subsec:Communication-Model},
the computing model in Section \ref{subsec:Computing-Model} and the
energy cost model in and \ref{subsec:Energy-Cost-Model}. Finally,
the problem formulation is proposed in Section \ref{subsec:Problem-Formulation}.
Besides, for clarity, the notations used in this work are listed in
Table \ref{tab:Notation-List}.

\begin{table}[t]
\centering\caption{\label{tab:Notation-List}Notation List}

\begin{tabular}{c|>{\centering}p{6cm}}
\hline 
\textbf{Notations} & \textbf{Parameters}\tabularnewline
\hline 
$\mathcal{I}$ & IoT user set, $i\in\mathcal{I}$.\tabularnewline
\hline 
$\mathcal{U}$ & UAV set, $u\in\mathcal{U}$.\tabularnewline
\hline 
$\mathcal{H}$ & HAP set, $h\in\mathcal{H}$.\tabularnewline
\hline 
$\sigma_{i}$ & Data size of IoT $i\in\mathcal{I}$.\tabularnewline
\hline 
$D_{i}$ & Maximum delay tolerated by IoT $i\in\mathcal{I}$.\tabularnewline
\hline 
$\rho{}_{u}$  & Computation resource cost of UAV $u$ to process $\textrm{1bit}$
data.\tabularnewline
\hline 
$\mu_{h}$  & Computation resource cost of HAP $h$ to process $\textrm{1bit}$
data.\tabularnewline
\hline 
$c_{iu}$ & Data rate of channel I2U.\tabularnewline
\hline 
$c_{uh}$ & Data rate of channel U2H.\tabularnewline
\hline 
$\mathbf{q}_{u}$ & Horizon location of UAV $u$.\tabularnewline
\hline 
$\mathbf{q}_{i}$ & Horizon location of IoT $i$.\tabularnewline
\hline 
$H_{u}$ &  Flight altitude of UAV $u$.\tabularnewline
\hline 
$N_{u}$ & The maximum number of IoT a UAV can serve.\tabularnewline
\hline 
$T_{iu}$ & Time cost to transmit the data of IoT $i$ to UAV $u$.\tabularnewline
\hline 
$T_{uh}$ & Time cost to transmit the data to HAP $h$ by UAV $u$.\tabularnewline
\hline 
$T_{u}^{i}$ & Time cost by UAV to complete the computation for IoT $i$.\tabularnewline
\hline 
$T_{h}^{i}$ &  Time cost by HAP $h$ to complete the computation for IoT $i$. \tabularnewline
\hline 
$P_{i}^{tr}$ & Transmission power of IoT $i$ to UAV $u$.\tabularnewline
\hline 
$P_{u}^{tr}$  & Transmission power of UAV $u$ to HAP $h$.\tabularnewline
\hline 
 $\varsigma_{u}$  & Energy consumption coefficient of UAV based computation.\tabularnewline
\hline 
 $\varsigma_{h}$  & Energy consumption coefficient of HAP based computation.\tabularnewline
\hline 
$E_{i}^{c}$ & Total energy cost of IoT $i$.\tabularnewline
\hline 
$E_{i}^{o}$ & Basic operation energy cost of IoT $i$.\tabularnewline
\hline 
$E_{i}^{tr}$ & Energy cost for data transmission from IoT $i$ to UAV $u$.\tabularnewline
\hline 
$E_{i}$  &  Energy budget of IoT .\tabularnewline
\hline 
$E_{u}^{c}$ & Total energy cost of UAV $u$.\tabularnewline
\hline 
$E_{u}^{o}$ & Basic energy operation cost of UAV $u$.\tabularnewline
\hline 
$E_{u}^{co}$ & Energy cost for computation of UAV $u$.\tabularnewline
\hline 
$E_{u}^{tr}$ & Energy cost for data transmission from UAV $u$ to HAP $h$ .\tabularnewline
\hline 
$E_{u}$  & Energy budget of UAVs.\tabularnewline
\hline 
$E_{h}^{c}$ & Total energy cost of HAP $h$.\tabularnewline
\hline 
$E_{h}^{o}$  & Basic operation cost of HAP $h$.\tabularnewline
\hline 
$E_{h}^{co}$ & Energy cost of HAP $h$ for computation.\tabularnewline
\hline 
$E_{h}$ & Energy budget of HAPs.\tabularnewline
\hline 
$C_{u}$ & Computing capability of UAV $u$.\tabularnewline
\hline 
$C_{h}$ & Computing capability of HAP $h$.\tabularnewline
\hline 
$\mathcal{M}_{1}$ & Matching in Algorithm \ref{alg:matching-IoT-UAV}.\tabularnewline
\hline 
 $\mathcal{M}_{2}$ & Matching in Algorithm \ref{alg:Externality-elimination}.\tabularnewline
\hline 
 & \textbf{Decision Variables}\tabularnewline
\hline 
$x_{u}^{i}$ & $x_{u}^{i}\in\left\{ 0,1\right\} $ indicates whether the task of
IoT $i\in\mathcal{I}$ is offloaded to UAV $u$.\tabularnewline
\hline 
$\beta_{u}^{i}$ & $\beta_{u}^{i}\in\left\{ 0,1\right\} $ indicates whether the task
of IoT $i\in\mathcal{I}$ is computed by UAV $u$.\tabularnewline
\hline 
$y_{h}^{i,u}$ & $y_{h}^{i,u}\in\left\{ 0,1\right\} $ indicates whether the task from
IoT $i$ is forwarded to HAP $h$ by UAV $u$. \tabularnewline
\hline 
$\gamma_{h}^{i}$ & $\gamma_{u}^{i}\in\left\{ 0,1\right\} $ indicates whether the task
of IoT $i\in\mathcal{I}$ is computed by HAP $h$.\tabularnewline
\hline 
\end{tabular}
\end{table}

\subsection{\textcolor{black}{Hierarchical} Aerial Computing Scenario \label{subsec:Hierarchical-scenario}}

As shown in Fig. \ref{fig:scenario}, the hierarchical aerial computing
framework is composed of UAVs and HAPs in the air, and terrestrial
IoT users in various applications, e.g., smart wearable devices, surveillance
cameras, smart framing, and IoT in disasters. \textcolor{black}{Note
that only the rotary-wing UAV is considered in the scenario, which
is able to float at a quasi-static position for a couple of hours.
Besides, HAPs serve as stable base stations in the air. Hence, the
hierarchical aerial computing model in the work is deemed as quasi-static.
}Both UAVs and HAPs are equipped with edge servers, and HAPs have
\textcolor{black}{stronger load capacity} than UAVs. The ground IoT
users have various computing demands, but with limited computing capability,
especially for the small size IoT device. As for the \textcolor{black}{lightweight
computation} demands, IoT devices can complete computing locally.
However, due to the limited computing and energy resources of IoT
devices, the computation-intensive demands may not be completed locally
by the IoT devices, and UAVs equipped with edge servers can provide
the computing service for these IoT devices via data offloading. Furthermore,
the payload for computation of UAV is limited, the computing tasks
on the UAV may fail. In this case, HAPs with \textcolor{black}{stronger}
payload can assist UAVs to accomplish the computation task from IoT
devices. In such a way, the UAV serves as a relay for the data from
IoT offloading to the HAP, rather than computation on the UAV. \textcolor{black}{Besides,
only binary computation offloading is considered in this model, }i.e.,
the computing task has two choices\footnote{The local computing by the IoT device itself is omitted in the model,
since local computing does not participate the offloading decisions.}: offloading to a UAV and computed by the edge server of the UAV,
or offloading to the HAP and computed by the edge server of the HAP,
according to the resource provision, as depicted in Fig. \ref{fig:scenario}. 

\subsection{Communication Model \label{subsec:Communication-Model}}

\subsubsection{Channel Model from IoT to UAV (I2U)}

\begin{figure}[t]
\centering

\includegraphics[scale=0.5]{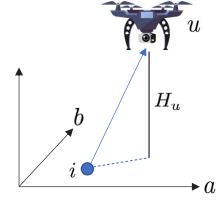}

\caption{Relative location of UAV and IoT.\label{fig:Relative-location}}
\end{figure}

\textcolor{black}{To avoid congestions, the orthogonal frequency division
is applied for the I2U channel, and the channel from IoT devices to
UAVs is} line-of-sight \cite{704-access-sky-B5g,1021-UAV-MEC-dongchao}.
Following \cite{yuye-TVT-UAV-MEC,jzy-IoTJ2020}, the channel gain
between IoT $i$ and UAV $u$ is \vspace{-4mm}

\begin{align}
G_{iu} & =\frac{G_{0}}{d_{iu}^{2}}=\frac{G_{0}}{\left(a_{u}-a_{i}\right)^{2}+\left(b_{u}-b_{i}\right)^{2}+H_{u}^{2}}\nonumber \\
 & =\frac{G_{0}}{\parallel\mathbf{q}_{u}-\mathbf{q}_{i}\parallel^{2}+H_{u}^{2}},\forall i\in\mathcal{I},u\in\mathcal{U},
\end{align}
where $d_{iu}$ is indicates the distance between IoT $i$ and UAV
$u$, and $G_{0}$ denotes the reference I2U channel gain at $d_{iu}=1\textrm{m}$.
As shown in Fig. \ref{fig:Relative-location}, $\mathbf{q}_{u}=\left\{ a_{u},b_{u}\right\} $
and $\mathbf{q}_{i}=\left\{ a_{i},b_{i}\right\} $ denote the horizon
location of UAV $u$ and IoT $i$, respectively. $H_{u}$ is the flight
altitude of UAV $u$. Then, the available data rate of the channel
from IoT $i$ to UAV $u$ is calculated as \vspace{-2mm}

\begin{align}
c_{iu} & =B_{iu}\cdot\textrm{log}_{2}\left(1+\frac{P_{i}^{tr}G_{iu}}{\delta^{2}}\right)\nonumber \\
 & =B_{iu}\cdot\textrm{log}_{2}\left(1+\frac{P_{i}^{tr}\iota_{0}}{\parallel\mathbf{q}_{u}-\mathbf{q}_{i}\parallel^{2}+H_{u}^{2}}\right),\forall i\in\mathcal{I},u\in\mathcal{U},
\end{align}
where $B_{iu}$ denotes the bandwidth of I2U channel, $\iota_{0}=\frac{G_{0}}{\delta^{2}}$
indicates the reference signal-to-noise ratio.\textcolor{black}{{} Recall
that $G_{iu}$ is the channel gain between IoT $i$ and UAV $u$.
}Hence, the time cost to transmit the data of IoT $i$ to UAV $u$
is 

\begin{equation}
T_{iu}=\frac{\sigma_{i}x_{u}^{i}}{c_{iu}},\forall i\in\mathcal{I},u\in\mathcal{U},
\end{equation}
in which binary variable $x_{u}^{i}$ indicates whether the task of
IoT $i$ is offloaded to UAV $u$, i.e.,\vspace{-2mm}

\[
x_{u}^{i}=\begin{cases}
1, & \textrm{task\;of\;IoT}\;i\;\textrm{is\;offloaded\;to\;UAV }u,\\
0, & \textrm{otherwise,}
\end{cases}
\]
and $\sigma_{i}$ is the data size of IoT $i$.

\subsubsection{Channel Model from UAV to HAP (U2H)}

According to \cite{jzy-IoTJ2020} and the Shannon theory, the achievable
data rate of U2H channel is\vspace{-3mm}

\begin{equation}
c_{uh}=B_{uh}\cdot\textrm{log}_{2}\left(1+\frac{P_{u}^{tr}G_{uh}L_{s}L_{l}}{k_{B}T_{s}B_{uh}}\right),\forall u\in\mathcal{U},h\in\mathcal{H},
\end{equation}
where $B_{uh}$ is the bandwidth of U2H channel, $G_{uh}$ is the
antenna power gain, $L_{l}$ is the total line loss, and $L_{s}=\left(\frac{c}{4\pi d_{uh}f_{uh}}\right)^{2}$
is the free space loss. Wherein, $c$ is the speed of light, $d_{uh}$
is the distance between UAV $u$ and HAP $h$, and $f_{uh}$ is center
frequency. $k_{B}$ is the Boltzmann's constant, and $T_{s}$ denotes
the system noise temperature. Besides, due to the long distance between
a UAV and a HAP, $d_{uh}$ is deemed as the perpendicular distance
between UAV $u$ and HAP $h$.\textcolor{black}{{} Note that to avoid
congestions, the orthogonal frequency division is also applied for
the U2H channel.}

Hence, the time cost to transmit the data of IoT $i$ to HAP $h$
from UAV $u$ can be calculated as

\begin{equation}
T_{uh}=\frac{\sigma_{i}y_{h}^{i,u}}{c_{uh}},\forall u\in\mathcal{U},h\in\mathcal{H},
\end{equation}
where $y_{h}^{i,u}\in\left\{ 0,1\right\} $ indicates whether the
task from IoT $i$ is forwarded to HAP $h$ by UAV $u$, i.e., \vspace{-2mm}

\[
y_{h}^{i,u}=\begin{cases}
1, & \textrm{data\;of\;IoT}\;i\;\textrm{is\;forwarded to HAP }h\textrm{ by UAV }u,\\
0, & \textrm{otherwise}.
\end{cases}
\]

\subsection{Computing Model \label{subsec:Computing-Model}}

For an IoT user, the computing demand can be offloaded to a UAV and
complete the computation on the UAV, or relayed by a UAV to a HAP
and complete the computation by the HAP \cite{1091-caching-MEC-HAP}.\textcolor{blue}{{} }

\subsubsection{UAV-based Computing}

In light of \cite{1088-UAV-MEC-timevaring}, denote $\rho{}_{u}$
as the computing resource consumed on UAVs to handle $\textrm{1bit}$
IoT data, i.e., the CPU cycles. Thus, the time cost by UAV to complete
the computation for IoT $i$ is\vspace{-3mm}

\begin{equation}
T_{u}^{i}=\frac{\sigma_{i}\beta_{u}^{i}}{C_{u}/\rho{}_{u}}=\frac{\sigma_{i}\beta_{u}^{i}\rho{}_{u}}{C_{u}},\forall i\in\mathcal{I},u\in\mathcal{U},
\end{equation}
where $C_{u}$ denotes the computation capability of UAV $u$, and
$\beta_{u}^{i}$ is the binary variable denoting whether the task
of IoT $i\in\mathcal{I}$ is computed by UAV $u$, in detail,\vspace{-2mm}

\[
\beta_{u}^{i}=\begin{cases}
1, & \textrm{task\;of\;IoT}\;i\;\textrm{is\;computed\;by\;UAV }u,\\
0, & \textrm{otherwise.}
\end{cases}
\]

\subsubsection{HAP-based Computing}

If the remaining computing resource of UAV cannot afford the IoT computing
task, the task will be offloaded to the HAP relayed by the UAV. Let
$\mu_{h}$ denote the computing resource cost of HAP $h$ to process
$\textrm{1bit}$ IoT data, and $C_{h}$ indicates the computation
capacity of HAP $h$. Accordingly, the time cost to complete the computation
for IoT $i$ by HAP $h$ is calculated as

\begin{equation}
T_{h}^{i}=\frac{\sigma_{i}\gamma_{h}^{i}}{C_{h}/\mu_{h}}=\frac{\sigma_{i}\gamma_{h}^{i}\mu_{h}}{C_{h}},\forall i\in\mathcal{I},h\in\mathcal{H},
\end{equation}
in which binary variable $\gamma_{u}^{i}\in\left\{ 0,1\right\} $
indicates whether the task of IoT $i\in\mathcal{I}$ is computed by
HAP $h$, \vspace{-2mm}

\[
\gamma_{u}^{i}=\begin{cases}
1, & \textrm{task\;of\;IoT}\;i\;\textrm{is\;computed by HAP }h,\\
0, & \textrm{otherwise.}
\end{cases}
\]

As above, the total time cost for IoT $i$ to complete necessary transmission
and computation is derived as\vspace{-2mm}
\begin{align}
T_{i}= & \underset{u\in\mathcal{U}}{\sum}\left(T_{iu}+T_{u}^{i}+\underset{h\in\mathcal{H}}{\sum}T_{uh}\right)+\underset{h\in\mathcal{H}}{\sum}T_{h}^{i}\nonumber \\
\!=\! & \underset{u\in\mathcal{U}}{\sum}\!\!\left(\!\!\frac{\sigma_{i}x_{u}^{i}}{c_{iu}}\!+\!\frac{\sigma_{i}\beta_{u}^{i}\rho{}_{u}}{C_{u}}\!+\!\underset{h\in\mathcal{H}}{\sum}\frac{\sigma_{i}y_{h}^{i,u}}{c_{uh}}\right)\!\!+\!\!\underset{h\in\mathcal{H}}{\sum}\frac{\sigma_{i}\gamma_{h}^{i}\mu_{h}}{C_{h}},\!\forall i\!\in\!\mathcal{I}.
\end{align}
Note that the delay to complete computation for IoT $i$ is related
with the time cost of transmission and computation processing. \textcolor{black}{Besides,
due to the small data size of the computation result, the delay as
well as the energy cost of computing result transmission are omitted
\cite{1067-SAG-MEC,1080-MEC-NOMA}.}

\subsection{Energy Cost Model\label{subsec:Energy-Cost-Model}}

\subsubsection{Energy Cost of IoT}

The energy cost $E_{i}^{c}$ of IoT $i$ is mainly composed by the
basic operation cost $E_{i}^{o}$ and the transmission cost $E_{i}^{tr}$,\vspace{-2mm}

\begin{align}
E_{i}^{c}= & E_{i}^{o}+E_{i}^{tr}=E_{i}^{o}+\underset{u\in\mathcal{U}}{\sum}P_{i}^{tr}T_{iu}\nonumber \\
= & E_{i}^{o}+\underset{u\in\mathcal{U}}{\sum}\frac{P_{i}^{tr}\sigma_{i}x_{u}^{i}}{c_{iu}},\forall i\in\mathcal{I},u\in\mathcal{U},
\end{align}
in which $P_{i}^{tr}$ denotes the transmission power from IoT $i$
to UAV $u$.

\subsubsection{Energy Cost of UAV}

The total energy cost $E_{u}^{c}$ of UAV $u$ is comprised of the
basic operation energy cost $E_{u}^{o}$, e.g. UAV hovering, the energy
cost $E_{u}^{co}$ for computation, and the transmission energy cost
$E_{u}^{tr}$. More concretely,\vspace{-2mm}

\begin{align}
E_{u}^{c}= & E_{u}^{o}+E_{u}^{co}+E_{u}^{tr}=E_{u}^{o}+\underset{i\in\mathcal{I}}{\sum}\varsigma_{u}C_{u}^{3}T_{u}^{i}+\underset{i\in\mathcal{I}}{\sum}\underset{h\in\mathcal{H}}{\sum}P_{u}^{tr}T_{uh}\nonumber \\
= & E_{u}^{o}+\underset{i\in\mathcal{I}}{\sum}\varsigma_{u}C_{u}^{2}\sigma_{i}\rho{}_{u}\beta_{u}^{i}+\underset{i\in\mathcal{I}}{\sum}\underset{h\in\mathcal{H}}{\sum}\frac{P_{u}^{tr}\sigma_{i}y_{h}^{i,u}}{c_{uh}},\forall u\in\mathcal{U},\label{eq:energycostUAV}
\end{align}
where $\varsigma_{u}$ denotes the energy consumption coefficient
depending on the chip structure of UAV's processor \cite{1088-UAV-MEC-timevaring}.
$P_{u}^{tr}$ is the power for UAV-based transmission to the HAP.

\subsubsection{Energy Cost of HAP}

The total energy cost $E_{h}^{to}$ of HAP $h$ is composed of basic
operation cost $E_{h}^{o}$ and the energy cost $E_{h}^{c}$ for computation,
\vspace{-2mm}

\begin{align}
E_{h}^{c}= & E_{h}^{o}+E_{h}^{c}=E_{i}^{o}+\underset{u\in\mathcal{U}}{\sum}P_{i}^{tr}T_{h}^{i}\nonumber \\
= & E_{h}^{o}+\underset{i\in\mathcal{I}}{\sum}\varsigma_{u}C_{h}^{2}\sigma_{i}\gamma_{h}^{i}\mu_{h},\forall h\in\mathcal{H},\label{eq:energycostHAP}
\end{align}
where $\varsigma_{u}$ is the energy consumption coefficient depending
on the chip structure of the HAP's processor. 

\subsection{Problem Formulation \label{subsec:Problem-Formulation}}

The objective is addressed to maximize the total IoT data computed
by the hierarchical aerial computing platforms (UAVs and HAPs), and
restricted by multiple resource and offloading decision constraints,
\vspace{-2mm}

\begin{align}
(\textrm{P0}):\underset{\boldsymbol{x,\beta,y,\gamma}}{\textrm{max}} & \;\underset{i\in\mathcal{I}}{\sum}\underset{u\in\mathcal{U}}{\sum}\underset{h\in\mathcal{H}}{\sum}\sigma_{i}\left(\beta_{u}^{i}+\gamma_{h}^{i}\right)\nonumber \\
\textrm{s.t.}\, & \underset{u\in\mathcal{U}}{\sum}x_{u}^{i}\leq1,\forall i\in\mathcal{I},\label{cons:IoT2UAV-01}\\
 & \beta_{u}^{i}+y_{h}^{i,u}=x_{u}^{i},\forall i\in\mathcal{I},u\in\mathcal{U},h\in\mathcal{H},\label{cons:UAV-flowconservation}\\
 & {\color{blue}{\color{black}\underset{i\in\mathcal{I}}{\sum}x_{u}^{i}}{\color{black}\leq N_{u},\forall u\in\mathcal{U},}}\label{cons:UAV-quota-IoT}\\
 & \gamma_{h}^{i}\leq\underset{u\in\mathcal{U}}{\sum}y_{h}^{i,u},\forall i\in\mathcal{I},h\in\mathcal{H},\label{cons:HAP-01}\\
 & E_{i}^{c}\leq E_{i},\forall i\in\mathcal{I}.\label{cons:IoT-EnergyCap}\\
 & E_{u}^{c}\leq E_{u},\forall u\in\mathcal{U},\label{cons:UAV-EnergyCap}\\
 & E_{h}^{c}\leq E_{h},\forall h\in\mathcal{H},\label{cons:HAP-EnergyCap}\\
 & T_{i}\leq D_{i},\forall i\in\mathcal{I},\label{cons:IoT-delay-cons}\\
 & x_{u}^{i}\in\left\{ 0,1\right\} ,\forall i\in\mathcal{I},u\in\mathcal{U},\\
 & \beta_{u}^{i}\in\left\{ 0,1\right\} ,\forall i\in\mathcal{I},u\in\mathcal{U},\\
 & y_{h}^{i,u}\in\left\{ 0,1\right\} ,\forall i\in\mathcal{I},u\in\mathcal{U},h\in\mathcal{H},\\
 & \gamma_{h}^{i}\in\left\{ 0,1\right\} ,\forall i\in\mathcal{I},h\in\mathcal{H},
\end{align}
where we have $\boldsymbol{x}=\{x_{u}^{i},\forall i\in\mathcal{I},u\in\mathcal{U}\}$,
$\boldsymbol{\beta}=\{\beta_{u}^{i},\forall i\in\mathcal{I},u\in\mathcal{U}\}$,
$\boldsymbol{y}=\{y_{h}^{i,u},\forall i\in\mathcal{I},u\in\mathcal{U},h\in\mathcal{H}\}$,
and $\boldsymbol{\gamma}=\left\{ \gamma_{h}^{i},\forall i\in\mathcal{I},h\in\mathcal{H}\right\} $,
denoting the variable vectors of IoT data offloading to the UAV, UAV-based
MEC, IoT data offloading to the HAP, and HAP-based MEC, respectively.
In P0, constraint (\ref{cons:IoT2UAV-01}) denotes that each IoT can
only connect to at most one UAV\textcolor{black}{. Note that not all
IoT data can be successfully offloaded to a UAV due to the resource
limitation.}\textcolor{blue}{{} }Constraint (\ref{cons:UAV-flowconservation})
implies the data flow conservation at a UAV. Constraint (\ref{cons:UAV-quota-IoT})
refers to the quota restriction of the UAV, i.e., the accommodated
IoT devices by a UAV cannot exceed the quota\textcolor{blue}{{} ${\color{black}N}_{{\color{black}u}}$}.
Constraint (\ref{cons:HAP-01}) depicts the coupled relation between
$\gamma_{h}^{i}$ and $y_{h}^{i,u}$. Constraints (\ref{cons:IoT-EnergyCap})-(\ref{cons:HAP-EnergyCap})
denote the energy capacity restrictions, and $E_{i}$, $E_{u}$, $E_{h}$
are the energy budget of IoT, UAV, and HAP, respectively. Constraint
(\ref{cons:IoT-delay-cons}) enforces the total time cost cannot exceed
the maximum tolerant delay $D_{i}$ of IoT $i$. 

It is observed that P0 is an integer programming problem, and is intractable
to solve especially in the case of large scale networks. Since the
complexity of exhaustive searching is related with the number of decision
variables of P0, i.e., $\mathcal{O}(2^{\mid\mathcal{I}\mid\cdot(2\mid\mathcal{U}\mid+\mid\mathcal{U}\mid\cdot\mid\mathcal{H}\mid+\mid\mathcal{U}\mid)})$,
and the various constraints further aggravate the complexity. Therefore,
efficient algorithms will be designed to deal with the complicated
problem in the following section.

\section{Algorithm Design\label{sec:Algorithm-Design}}

As the above discussion, P0 is in the form of integer programming,
which is intractable to directly obtain the solution. In this section,
we adopt the matching game based mechanisms to handle the offloading
decision from IoT devices to UAVs in Section \ref{subsec:Matching-based-Algorithm}.\textcolor{black}{{}
Further, to} eliminate the external effect among different IoT devices,
the externality elimination algorithm is presented in Section \ref{subsec:Eliminating-the-Externality}.
As for the data offloading from UAVs to HAPs, a heuristic algorithm
is designed in Section \ref{subsec:Data-Offloading-uav-hap}. Besides,
an adjustment algorithm is proposed to take full advantage of aerial
resources in Section \ref{subsec:Adjustment-Algorithm}.

\subsection{Matching based Algorithm for IoT Data Offloading to UAV\label{subsec:Matching-based-Algorithm}}

\subsubsection{Preliminary of Matching Game Theory}

As a Nobel Prize winning mechanism in Economic Science, matching game
theory can handle the social and marketing problems in a distributed
mode \cite{BOOK-matching}. Besides, matching game theory finds wide
applications in network management \cite{929-TMC2020-Neetu,541-RVV-GuYunan}\textcolor{black}{.
The primary advantage of matching game theory is that it considers
the preference of the participated agents, and provides the distributed
solutions. The common thread in the matching game theory is to find
a stable matching for the participated agents with special preference
over another set of agents.}

\subsubsection{\textcolor{black}{Matching between IoT and UAV}}

\textcolor{black}{Inspired by the matching game theory, the offloading
problem from IoT devices to UAVs in P0 can be deemed as a matching
problem with two sets of agent: IoT devices and UAVs, and the problem
is a two-sided matching. Besides, the constraints in P0 can be implied
in the preference lists of IoT devices and UAVs, respectively. }Since
an IoT user can only connect to one UAV, and each UAV can serve a
couple of IoT users, the matching between IoT devices and UAVs is
in the many-to-one form. 

Primarily, the preference list of IoT devices on UAVs is defined as
\vspace{-2mm}

\begin{equation}
PL_{i}=\lambda_{1}C_{u}^{r}+\lambda_{2}E_{u}^{r}+\lambda_{3}c_{iu},\label{eq:Preference-IoT}
\end{equation}
since IoT devices prefer the UAV with larger residual computing capacity
$C_{u}^{r}$, residual energy budget $E_{u}^{r}$, and available channel
capacity $c_{iu}$, and $\lambda_{1}$, $\lambda_{2}$, and $\lambda_{3}$
denote the weighted parameters. For example, IoT $i$ prefers UAVs
$u^{1}$ to $u^{2}$ if UAV $u^{1}$ has priority over UAV $u^{2}$
in $PL_{i}$, and it is expressed as $u^{1}\succ_{i}u^{2}$. In this
case, IoT $i$ will choose UAV $u^{1}$ in matching $\mathcal{M}_{1}$,
represented as $\mathcal{M}_{1}(i)=u^{1}$.

The preference list of UAVs on IoT devices is expressed as \vspace{-3mm}

\begin{equation}
PL_{u}=\iota_{1}\sigma_{i}+\iota_{2}D_{i}.\label{eq:Preference-UAV}
\end{equation}
Recall that $\sigma_{i}$ is the data size of IoT $i$ and $D_{i}$
is the maximum delay tolerated by IoT $i$, and so UAVs prefer the
IoT devices with a large data size and tolerant delay, which is in
accordance with the objective of P0. $\iota_{1}$ and $\iota_{2}$
indicate the the weighted parameters. For example, if there exist
two IoT devices $i^{1}$ and $i^{2}$ in $PL_{u}$, UAV $u$ will
select IoT $i^{2}$ as its partner, i.e., $\mathcal{M}_{1}(u)=i^{2}$,
if $i^{2}\succ_{u}i^{1}$ and UAV $u$ has only one vacancy; in another
hand, if UAV $u$ has more than one residual quota, both IoT devices
$i^{1}$ and $i^{2}$ can be matched to the UAV.
\begin{algorithm}[t]
\caption{\textcolor{black}{Many-to-one matching between IoT and UAV (MIU)\label{alg:matching-IoT-UAV}}}

\begin{algorithmic}[1]

\REQUIRE $\mathcal{I}$, $\mathcal{U}$, $\sigma_{i}$, and $D_{i}$.

\ENSURE Stable matching $\mathcal{M}_{1}$, and $\boldsymbol{x}$.

\STATE\textit{Initialization:}\textcolor{black}{{} Construct preference
lists $PL_{i}$ and $PL_{u}$. Set }$\mathcal{M}_{1}=\emptyset$,
and $flag=1$. 

\WHILE{$flag==1$}

\STATE $flag=0$.

\FOR{each unmatched IoT $i$}

\IF{$PL_{i}\neq\emptyset$}

\STATE Select the most preferred UAV $u\in PL_{i}$ as $\mathcal{M}_{1}(i)$. 

\IF{$|\mathcal{M}_{1}(i)|==N_{u}$}

\STATE Select the worst matched IoT $i'$ in $\mathcal{M}_{1}(i)$. 

\IF{$i\succ_{u}i'$}

\STATE Swap $i$ and $i'$ in $\mathcal{M}_{1}(i)$.

\ELSE

\STATE Delete $i$ from $PL_{i}$.

\ENDIF

\STATE Set $flag=1$. 

\ENDIF

\STATE Add pair $(i,u)$ to $\mathcal{M}_{1}$.\label{alg1:Add-pair}

\ENDIF

\ENDFOR

\ENDWHILE

\end{algorithmic}
\end{algorithm}

In particular, to indicate whether a matching arrives stable, the
blocking pair in\textcolor{black}{{} many-to-one matching} is defined
as:
\begin{defn}
\textbf{\textcolor{black}{(Blocking pair in many-to-one matching):
}}\textcolor{black}{In the many-to-one matching between IoT devices
and a UAV, a pair $(i,u)\notin\mathcal{M}_{1}$ is deemed as a blocking
pair for $\mathcal{M}_{1}$ if: 1) IoT $i$ is unserved or IoT $i$
prefers UAV $u$ to its current matching $\mathcal{M}_{1}(i)$; 2)
UAV $u$ is underutilized or prefers IoT $i$ to at least one existed
matching $\mathcal{M}_{1}(u)$ in $\mathcal{M}_{1}$. }
\end{defn}
Inspired by the Gale-Shapley mechanism \cite{557-GaleShapley}, the
many-to-one matching between IoT devices and a UAV is detailed in
Algorithm \ref{alg:matching-IoT-UAV}. To begin with, the preference
lists of IoT devices and UAVs are, respectively, constructed following
(\ref{eq:Preference-IoT}) and (\ref{eq:Preference-UAV}). \textcolor{black}{$\mathcal{M}_{1}$
is initialized as $\emptyset$ and $flag$ is applied to controlling
algorithm execution.} Then, the algorithm starts from IoT $i$ (IoT-oriented)
and it selects the most preferred UAV $u$ in $PL_{i}$ as its partner,
i.e., $u=\mathcal{M}_{1}(i)$, constructing a pair $(i,u)$. If the
selected UAV $u$ is undersubscribed, directly add $(i,u)$ to $\mathcal{M}_{1}$,
as the step \ref{alg1:Add-pair} of Algorithm \ref{alg:matching-IoT-UAV}.
Otherwise, if the quota of UAV $u$ is full, i.e., $|\mathcal{M}_{1}(i)|==N_{u}$,
IoT $i$ is compared with the worst matched IoT $i'$ in $\mathcal{M}_{1}$.
If IoT $i$ is superior to $i'$ in the UAV's preference list $PL_{u}$,
$(i,u)$ is a blocking pair for matching \textcolor{black}{$\mathcal{M}_{1}$,
and }$i$ and $i'$ are swapped for stable; or $i$ will be deleted
from $PL_{u}$. Note that the UAV's quota constraint (\ref{cons:UAV-quota-IoT})
in P0 is implied in Algorithm \ref{alg:matching-IoT-UAV}. A stable
matching $\mathcal{M}_{1}$ is obtained after the termination of Algorithm
\ref{alg:matching-IoT-UAV}, when there exists no blocking pair or
all preference lists of IoT are empty. The complexity of Algorithm
\ref{alg:matching-IoT-UAV} is related to the number of potential
IoT-UAV pairs \cite{541-RVV-GuYunan}, i.e., $\mathcal{O}(|\mathcal{I}|\cdot|\mathcal{U}|)$. 

Notice that if the preference lists\textcolor{black}{{} $PL_{i}$ and
$PL_{u}$ are fixed, the preference of any participant merely depends
on the certain information about the participants in another set.
However, from (\ref{eq:Preference-IoT}), we observe that the IoT's
preference is influenced by the choices of other participants, since
the residual computing resource and }energy budget of UAVs may change
with different matching decision, and the matching result $\mathcal{M}_{1}$
from Algorithm \ref{alg:matching-IoT-UAV} \textcolor{black}{may not
stable} incurred by varying preference of IoT.\textcolor{black}{{} Such
a matching with the interplay of different participants' preferences
is the matching with externality \cite{BOOK-Roth}. Notice here, the
external effect is caused by any matching with the inter-dependence
of the participants' preferences \cite{541-RVV-GuYunan}.} For example,
if a UAV is chosen by too many IoT devices, and in this case only
a small part of computation and energy of the UAV is allocated to
each IoT user, so some IoT devices may have the incentive to change
to a different UAV that has more available resources. Hence, the issue
of externality should be tackled for a final stable matching.

\subsection{Eliminating the Externality\label{subsec:Eliminating-the-Externality}}

As discussed above, the externality in the matching between IoT devices
and UAVs should be dealt with, and we propose the externality elimination
algorithm to re-stabilize the matching in Algorithm \ref{alg:Externality-elimination}.
Specifically, the invalid IoT-UAV pair, i.e., the IoT failed to be
matched with all UAVs by Algorithm \ref{alg:matching-IoT-UAV}, should
be removed from $P_{u}$ and UAVs' preference lists are updated. Then,
the matching will be re-stabilized, and at this point, the new strategy
focuses on how to improve the performance of IoT devices, since the
preference of IoT devices is affected by the externality. More concretely,
it becomes a problem in regard to one-side stability. Such stability
bases on the equilibrium among all IoT devices, and it is defined
as ``Pareto optimal'' as follows \cite{BOOK-matching}.

\begin{algorithm}[t]
\caption{\label{alg:Externality-elimination}Externality\textcolor{black}{{}
elimination algorithm (EEA)}}

\begin{algorithmic}[1]

\REQUIRE Matching result $\mathcal{M}_{1}$ from Algorithm \ref{alg:matching-IoT-UAV},
and the updated preference list $PL_{i}$ related with $\mathcal{M}_{1}$.

\ENSURE Re-stabilized pairwise-stable matching $\mathcal{M}_{2}$,
and $\boldsymbol{x}$.

\STATE\textit{ }$\mathcal{M}_{2}=\mathcal{M}_{1}$.

\STATE\textcolor{black}{{} Remove invalid (IoT, UAV) pairs related
IoT devices.}

\WHILE{$\mathcal{M}_{2}$ is not Pareto optimal}

\STATE Search the unstable (IoT, IoT) blocking pairs $BL$ in terms
of $PL_{i}$ . 

\FOR{each $(i,i')\in BL$}

\IF{$\exists i\in\mathcal{M}_{2}(u)\cup\mathcal{M}_{2}(u')$, $\bigtriangleup U(i)>0$}

\STATE $(i,i')$ are permitted to switch partners.

\ELSE

\STATE $(i,i')$ are not permitted to switch partners.

\ENDIF

\ENDFOR

\STATE Find the optimal blocking pair $(i^{*},i'^{*})$.

\STATE $i^{*}$ and $i'^{*}$ exchange partners.\label{alg2:exchange-partner}

\STATE $\mathcal{M}_{2}=\mathcal{M}_{2}/\{(i^{*},\mathcal{M}_{2}(i^{*})),(i'^{*},\mathcal{M}_{2}(i'^{*}))\}$.\label{alg2:matching-update1}

\STATE $\mathcal{M}_{2}=\mathcal{M}_{2}\cup\{(i^{*},\mathcal{M}_{2}(i'^{*})),(i'^{*},\mathcal{M}_{2}(i^{*}))\}$.\label{alg2:matching-update2}

\ENDWHILE

\end{algorithmic}
\end{algorithm}
\begin{defn}
\textbf{\textcolor{black}{(Pareto optimal): }}\textcolor{black}{A
matching $\mathcal{M}$ is in Pareto optimal, if there exists no other
matching $\mathcal{M}'$ so that some IoT are better off in $\mathcal{M}'$
and no IoT is worse off.}
\end{defn}
Accordingly, the definition of the blocking pair in one-sided matching
is expressed as:
\begin{defn}
\textbf{\textcolor{black}{(Blocking pair in one-sided matching): }}\textcolor{black}{An
IoT pair $(i,i')$ is a blocking pair in the one-sided matching if
both IoT devices $i$ and $i'$ can be better off if they swap with
their matched UAVs.}
\end{defn}
The eternality\textcolor{black}{{} elimination algorithm is detailed
in Algorithm }\ref{alg:Externality-elimination}. Specifically, the
matching result $\mathcal{M}_{1}$ from Algorithm \ref{alg:matching-IoT-UAV}
and the updated preference list $PL_{i}$ related with $\mathcal{M}_{1}$
are regarded as the input for \textcolor{black}{Algorithm }\ref{alg:Externality-elimination},
and finally a re-stabilized matching $\mathcal{M}_{2}$ is acquired.
Firstly, the invalid (IoT, UAV) pairs related IoTs are removed. Then,
the Pareto optimality of matching $\mathcal{M}_{2}$ is checked and
the unstable (IoT, IoT) blocking pairs are found. In addition, $\mathcal{M}_{2}(i)=u$
and $\mathcal{M}_{2}(i')=u'$ denote that in matching $\mathcal{M}_{2}$,
IoT $i$ and UAV $u$ are paired while IoT $i'$ and UAV $u'$ are
paired. Define the utility of IoT $i$ as \vspace{-2mm}

\begin{equation}
U(i)=\lambda_{1}C_{u}^{r}+\lambda_{2}E_{u}^{r}+\lambda_{3}c_{iu},
\end{equation}
 and \vspace{-3mm}

\begin{equation}
\bigtriangleup U(i)=U(i)'-U(i),
\end{equation}
in which $U(i)'$ refers to the utility of IoT $i$ after switching
partner with IoT $i'$. Accordingly, the optimal blocking pair is
expressed as\vspace{-3mm}

\begin{equation}
(i^{*},i'^{*})=\underset{(i,i')}{\textrm{argmax}}\left(\underset{i\in\mathcal{M}_{2}(u)}{\sum}\!\!\!\!\bigtriangleup U(i)+\!\!\!\!\!\underset{i'\in\mathcal{M}_{2}(u')}{\sum}\!\!\!\!\bigtriangleup U(i')\right),
\end{equation}
and the IoT pair $(i^{*},i'^{*})$ is permitted to exchange partners
as step \ref{alg2:exchange-partner} of \textcolor{black}{Algorithm
}\ref{alg:Externality-elimination}. After that, matching $\mathcal{M}_{2}$
is updated as the steps \ref{alg2:matching-update1} and \ref{alg2:matching-update2}
in \textcolor{black}{Algorithm }\ref{alg:Externality-elimination}.
Due to the irreversibility of each switch and the limited number of
IoT pairs, \textcolor{black}{the convergence of Algorithm }\ref{alg:Externality-elimination}
is guaranteed. Actually, since the realization of Algorithm \ref{alg:Externality-elimination}
relies on iteratively searching the best blocking pair and exchanging
their partners, and the key is to find all blocking pairs, which require
traversing the preferences lists of all IoT devices. The number of
comparing operation is related to $|\mathcal{I}|\cdot|\mathcal{U}|$.
Besides, the number of iterations to search and swap blocking pairs
are related to $|\mathcal{I}|\cdot|\mathcal{I}|$. In the worst case,
the termination for Algorithm \ref{alg:Externality-elimination} has
a time complexity of $\mathcal{O}(|\mathcal{I}|\cdot|\mathcal{U}|\cdot|\mathcal{I}|\cdot|\mathcal{I}|)$,
i.e., $\mathcal{O}(|\mathcal{I}|^{3}\cdot|\mathcal{U}|)$. In practice,
the time complexity is lower than the theoretical analysis.

\begin{algorithm}[t]
\caption{\label{alg:UAV-HAP}Heuristic Algorithm for data offloading from UAV
to HAP (HA)}

\begin{algorithmic}[1]

\REQUIRE $\mathcal{M}_{2}$, $\sigma_{i}$ and $D_{i}$ of IoT in
$\mathcal{M}_{2}$, and $\boldsymbol{x}$.

\ENSURE $\boldsymbol{\beta}$, $\boldsymbol{y}$ and $\boldsymbol{\gamma}$
of P0.

\STATE Initialize $\varOmega=\emptyset.$

\STATE \textsl{\textcolor{black}{Check the IoT data at UAVs:}}

\FOR{each UAV $u$}

\IF{$\exists$ matched IoT $i$, the delay requirement cannot be satisfied by UAV $u$}

\STATE Rank these IoT data according to $\iota_{1}\sigma_{i}+\iota_{1}D_{i}$
in a descending order, and add into $\varOmega$. 

\STATE Offload the first IoT's data in $\varOmega$ to HAP. \label{Alg3: offload-to-HAP}

\IF{the residual IoT data in $\varOmega$ can be satisfied by $u$}\label{alg3:check-Set}

\STATE Go to step \ref{Alg3:check-HAP}.

\ELSE

\STATE $\varOmega$ is updated by deleting the offloaded IoT data,
and go to step \ref{Alg3: offload-to-HAP}.

\ENDIF \label{alg3:end-Check-set}

\ENDIF

\ENDFOR

\STATE \textsl{\textcolor{black}{Check the IoT data at the HAP:}}\label{Alg3:check-HAP}

\IF{$\exists$ IoT whose the delay requirement cannot be satisfied by the HAP}

\STATE Delete the IoT with the smallest data size $\sigma_{i}$,
update the residual resources of the HAP, and go to step \ref{Alg3:check-HAP}.

\ELSE

\STATE Return $\boldsymbol{\beta}$, $\boldsymbol{y}$ and $\boldsymbol{\gamma}$.

\ENDIF

\end{algorithmic}
\end{algorithm}

\subsection{Data Offloading from UAV to HAP\label{subsec:Data-Offloading-uav-hap}}

After Algorithms \ref{alg:matching-IoT-UAV} and \ref{alg:Externality-elimination},
IoT devices are successfully matched with UAVs, and the data from
IoT devices can be offloaded to the matched UAVs. However, due to
the limited computation capacity and energy budget of UAVs, the delay
requirement of IoT devices may not be satisfied. Note that HAPs are
still unoccupied, and UAVs can offload some IoT data to HAPs to alleviate
the overload on UAVs, and satisfy the delay requirement of IoT devices.
To address this issue, we propose the \textcolor{black}{heuristic}
algorithm for data offloading from UAVs to HAPs, detailed in Algorithm
\ref{alg:UAV-HAP}, and we only consider one HAP in this work since
the multiple HAPs can be handled with tractability. The results from
Algorithm \ref{alg:Externality-elimination} act as the input of Algorithm
\ref{alg:UAV-HAP}, and an auxiliary parameter $\varOmega$ is set
as $\emptyset$. Then, the IoT data offloaded at UAVs are checked.
In particular, with regard to a UAV $u$, if there exist IoT data
on the UAV whose delay requirement cannot be satisfied, these IoT
data are ranked according to $\iota_{1}\sigma_{i}+\iota_{1}D_{i}$
in a descending order and added in set $\varOmega$. After that, the
first IoT data in $\varOmega$ is offloaded to the HAP. Afterwards,
the satisfaction of residual IoT in $\varOmega$ is further checked
from step \ref{alg3:check-Set} to step \ref{alg3:end-Check-set}
of Algorithm \ref{alg:UAV-HAP}, since the available resource of the
UAV increases after offloading data to the HAP. Then, the IoT data
offloaded at the HAP are checked: the IoT data whose delay requirement
cannot be satisfied are found and the IoT data with the smallest data
size $\sigma_{i}$ is deleted, updating the residual resources of
the HAP and continuing the iteration until all IoT data at the HAP
can be satisfied. In the end of Algorithm \ref{alg:UAV-HAP}, variables
$\boldsymbol{\beta}$, $\boldsymbol{y}$ and $\boldsymbol{\gamma}$
in P0 are obtained.

The complexity of Algorithm \ref{alg:UAV-HAP} is composed of two
parts: checking the IoT's data at the UAV and checking the IoT data
at the HAP. In the worst case, the complexity in the first stage is
related to $|\mathcal{U}|\cdot|\mathcal{I}|$, and the second stage
is incurred by $|\mathcal{H}|\cdot|\mathcal{I}|$. Accordingly, the
complexity of Algorithm \ref{alg:UAV-HAP} is $\mathcal{O}((|\mathcal{U}|+|\mathcal{H}|)\cdot|\mathcal{I}|)$.

\subsection{\textcolor{black}{Adjustment Algorithm\label{subsec:Adjustment-Algorithm}}}

Since a couple of IoT data are offloaded from UAVs to the HAP by Algorithm
\ref{alg:UAV-HAP}, UAVs may\textcolor{black}{{} own some }redundant
resources. In this case, if there exist \textcolor{black}{some} unserved
IoT devices after execution of Algorithm \ref{alg:UAV-HAP}, we further
propose Algorithm \ref{alg:adjustment} to take full advantage of
UAVs' resources. Firstly, initialize the assistant parameter $\varOmega$
as $\emptyset.$ Then, if there exist unserved IoT devices after Algorithm
\ref{alg:UAV-HAP}, these IoT devices are ranked in terms of $D_{i}/\sigma_{i}$
in a descending order in a set $\varOmega$, since the IoT devices
with large $D_{i}/\sigma_{i}$ have greater possibility to be served
by the UAV with residual resources. The first IoT device in $\varOmega$
has the advantage to offloading to UAVs, as shown in the steps \ref{alg4:checkUAV}
to \ref{alg4:endfor}. The iteration is terminated until $\varOmega=\emptyset$.
With regard to the complexity, it is incurred by the size of $\varOmega$,
at worst with $|\mathcal{I}|$, and the size of $\mathcal{U}$, i.e.,
$\mathcal{O}(|\mathcal{I}|\cdot|\mathcal{U}|).$ 

\begin{algorithm}[t]
\caption{\label{alg:adjustment}Algorithm for adjustment (AA)}

\begin{algorithmic}[1]

\STATE Initialize $\varOmega=\emptyset.$

\STATE \textsl{\textcolor{black}{Check if there are any unserved
IoT after Algorithm}} \ref{alg:UAV-HAP}\textsl{\textcolor{black}{.}}

\IF{$\exists$ unserved IoT devices}

\STATE Rank these IoT devices in terms of $D_{i}/\sigma_{i}$ in
a descending order in a set $\varOmega$. 

\WHILE{$\varOmega\neq\emptyset$}\label{Alg4:while}

\FOR{each UAV $u$} \label{alg4:checkUAV}

\IF{the UAV can accormodate the first IoT $i$ in $\varOmega$}

\STATE IoT $i$ is offloaded to the UAV. $\varOmega=\varOmega/i$.
Update residual resources of the UAV. Go to step \ref{Alg4:while}.

\ENDIF

\ENDFOR \label{alg4:endfor}

\STATE $\varOmega=\varOmega/i$. 

\ENDWHILE

\ENDIF

\end{algorithmic}
\end{algorithm}
\begin{figure}[t]
\centering

\includegraphics[scale=0.44]{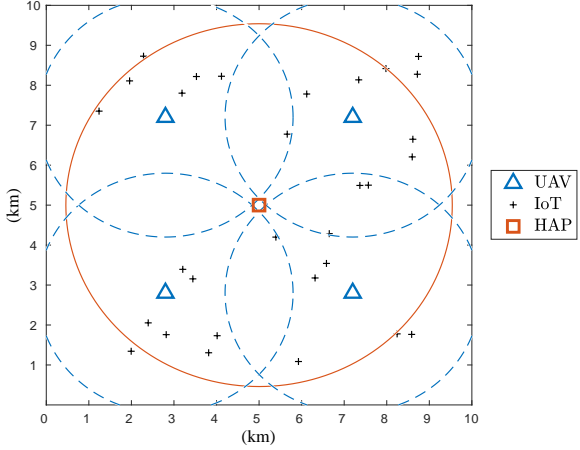}

\caption{Coverage of the UAV and HAP for IoT ($|\mathcal{I}|=30$).\label{simu:Coverage-example} }
\end{figure}

\section{Performance Evaluation \label{sec:Performance-Evaluation}}

In this section, we conduct simulations to evaluate the hierarchical
aerial computing mechanism and the proposed algorithms. The algorithm
design are implemented in MATLAB, and the optimization tools CVX as
well as MOSEK are also employed.
\begin{figure}[t]
\centering

\subfloat[Complexity. \label{sim:timeComplexity}]{\centering

\includegraphics[scale=0.44]{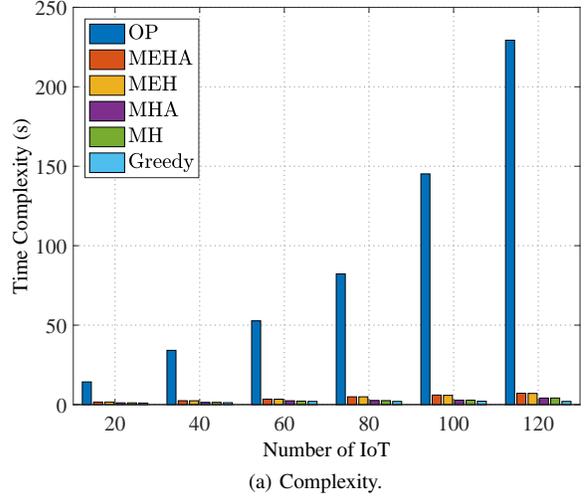}

}

\subfloat[Optimization results.\label{sim:optimizaitonResults}]{\centering

\includegraphics[scale=0.44]{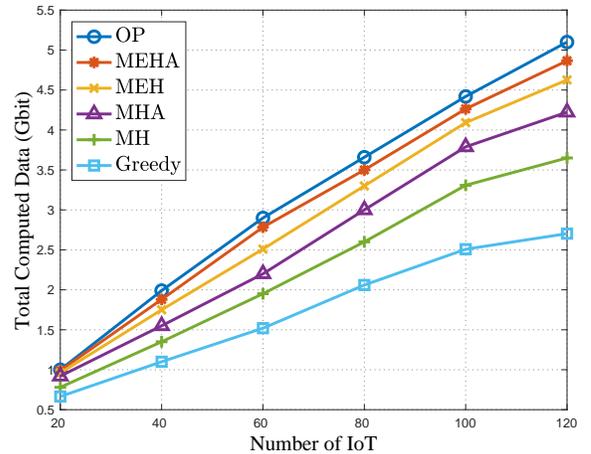}

}

\caption{Performance of proposed algorithms.\label{sim:Performance-alg}}
\end{figure}

\subsection{Simulation Setup}

\textcolor{black}{Simulations are conducted in the scenario: one HAP
with height of $20\textrm{km}$, 4 UAVs with altitude of $2\textrm{km}$
are uniformly distributed in the area with size of $10\textrm{km}\times10\textrm{km}$,
and terrestrial IoT users are randomly distributed in this area. An
illustration of the simulation scenario with respect to the coverage
relationships of the UAV and the HAP for IoT devices is shown in Fig.
\ref{simu:Coverage-example}. Note that UAVs are in the coverage of
the HAP, and terrestrial users are in the coverage of UAVs.} The data
size of IoT $\sigma_{i}$ is randomly generated from $[\textrm{10Mbit,100Mbit}]$,
and the maximum delay $D_{i}$ tolerated by IoT devices is randomly
generated in\textcolor{blue}{{} }\textcolor{black}{$[10\textrm{s},200\textrm{s}]$}
\cite{1064-UAV-MEC-IoT}. The quota of a UAV is set as $N_{u}=50$.
Besides, following \cite{1054-UAV-MEC-IoT,1052-UAV-MEC-energy,979-HAP-zhanglong,1091-caching-MEC-HAP},
the computation and communication related parameters are set as: $\rho_{u}=\textrm{270cycles/bit}$,
$\mu_{h}=\textrm{1100cycles/bit}$, $C_{u}=10^{9}\textrm{cycles/s}$,
$C_{h}=5\times10^{10}\textrm{cycles/s}$,\textcolor{black}{{} ${\color{blue}{\color{black}\varsigma_{u}=\varsigma_{h}=10^{-28}}}$,}
$B_{uh}=20\textrm{MHz}$, $G_{uh}=15\textrm{dB}$ $k_{B}=1.38\times10^{-23}\textrm{J/K}$,
$T_{s}=1000\textrm{K}$, and $f_{uh}=2.4\textrm{GHz}$. \textcolor{black}{The
power related parameters are set as $P_{i}^{tr}=0.5W$, $P_{u}^{tr}=10W$,
$E_{i}=100\textrm{J}$, $E_{u}=100\textrm{KJ}$, and $E_{h}=1000\textrm{KJ}$.
In addition, the} parameters in the matching based algorithm are set
as $\lambda_{1}=\lambda_{2}=0.4,$ $\lambda_{3}=0.2$, and $\iota_{1}=\iota_{2}=0.5$.
\begin{figure}[t]
\centering

\subfloat[Total computed data.\label{sim:computedData_mode}]{\centering

\includegraphics[scale=0.44]{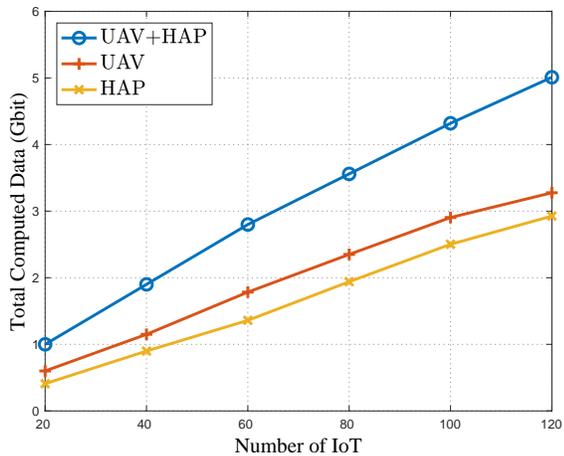}

}

\subfloat[Number of served IoT.\label{sim:served_IoT_mode}]{\centering

\includegraphics[scale=0.44]{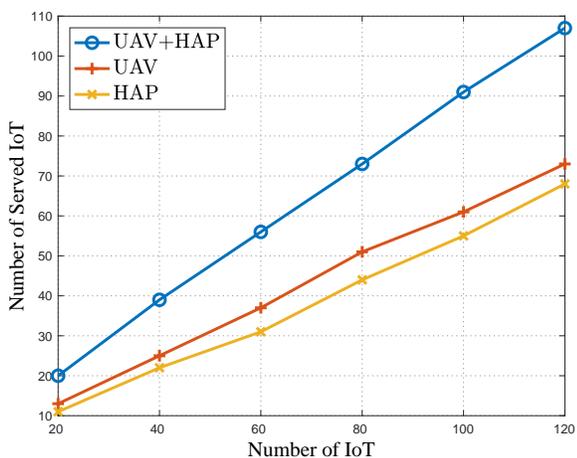}

}

\caption{Performance of different aerial computing mode. \label{sim:performance-comuptingMode}}
\end{figure}

\subsection{Performance Evaluation }

To evaluate the efficiency of the proposed algorithms, we compare
the combination of algorithms MIU+HA (MH), MIU+EEA+HA (MEH), MIU+HA+AA
(MHA), MIU+EEA+HA+AA (MEHA), as well as the optimal solution (OP)
obtained by the optimization tools, and the greedy offloading strategy.
Specifically, Fig. \ref{sim:Performance-alg} provides the performance
of the proposed algorithms, including the complexity in Fig. \ref{sim:timeComplexity}
and optimization results in Fig. \ref{sim:optimizaitonResults}. It
is observed that the MEHA can obtain the near optimal solution with
low complexity, compared with the optimal solution OP. The combined
algorithms MEH without adjustment of Algorithm \ref{alg:adjustment},
and MHA without the externality elimination by Algorithm \ref{alg:Externality-elimination}
perform worse than MEHA. Besides, the performance of the algorithm
MH without adjustment and externality elimination, as well as the
greedy strategy is undesirable, especially with a large number of
IoT. 
\begin{figure}[t]
\centering

\subfloat[Impact of HAP's computation capability ($C_{u}=10^{9}\textrm{cycles}$).\label{sim:Impact-of-HAP's computing}]{\centering

\includegraphics[scale=0.44]{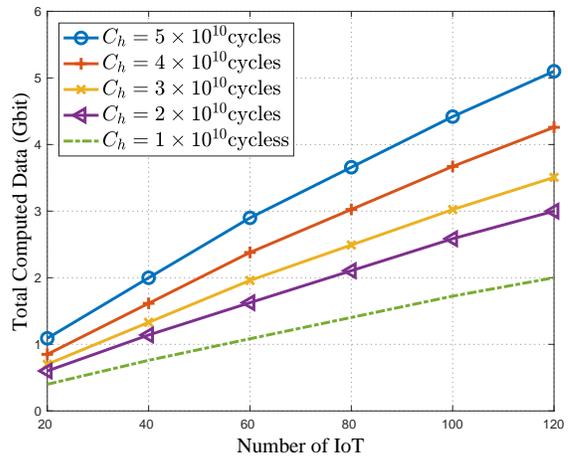}

}

\subfloat[Impact of UAV's computation capability ($C_{h}=5\times10^{10}\textrm{cycles}$).\label{sim:Impact-of-UAV's computing}]{\centering

\includegraphics[scale=0.44]{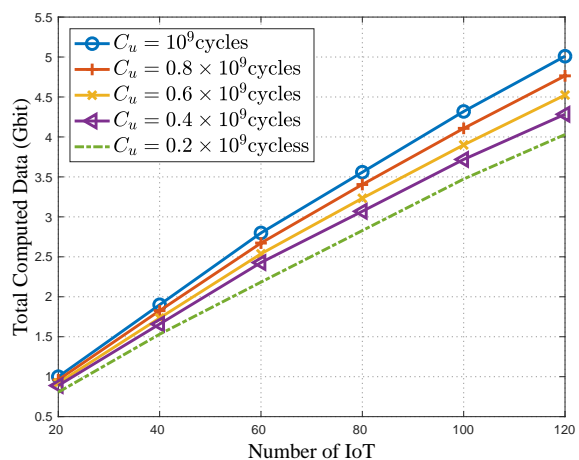}

}

\caption{Impact of computation capability on network performance.\label{sim:Impact-of-computing}}
\end{figure}

In Fig. \ref{sim:performance-comuptingMode}, we explore the performance
of different aerial computing mode, including UAV+HAP modes (hierarchical
computing mode we proposed in this work), UAV based aerial computing
mode (without HAP), HAP based aerial computing mode (without UAV).
In particular, algorithm MEHA is applied to the UAV+HAP mode, Algorithm
\ref{alg:matching-IoT-UAV} is used for the UAV mode, and the HAP
mode employs the many-to-one matching based strategy. It is observed
that the proposed hierarchical computing mode composed of UAVs and
HAPs performs better than both the UAV based aerial computing mode
and the HAP based aerial computing mode in terms of total computed
data and the number of served users. The reason that the UAV based
computing mode is better than the HAP based computing mode boils down
to the closer distance between IoT devices and UAVs, the limited IoT
transmission power, and there exist IoT devices out of the coverage
of the HAP. Hence, the advantage of cooperating UAVs and HAPs to provide
hierarchical aerial computing for IoT devices is verified. 

We further study the impacts of computation capacity of UAVs and HAPs
on the optimization performance in Fig. \ref{sim:Impact-of-computing}.
Specifically, in Fig. \ref{sim:Impact-of-HAP's computing}, the decrement
of HAP's computation capability imposes negative impact on the total
computed data, and there exists a similar effect from the UAV's computation
capability on the total computed data in Fig. \ref{sim:Impact-of-UAV's computing}.
It is noted that the variation of HAP's computation capacity has more
prominent impacts than the variation of UAV's computation capacity
on the optimization results, since the powerful computing capability
of the HAP provides computation service for a large number of IoT
devices. In fact, with the decrement of UAV's computing capacity,
the IoT data being computed at the UAV deceases and the UAV will have
more energy to relay the IoT data to the HAP, so the impact of UAV's
computation capacity variation is mild. 
\begin{figure}[t]
\centering

\subfloat[Impact of HAP's computation capability ($C_{u}=10^{9}\textrm{cycles}$).\label{sim:energycost-HAP}]{\centering

\includegraphics[scale=0.44]{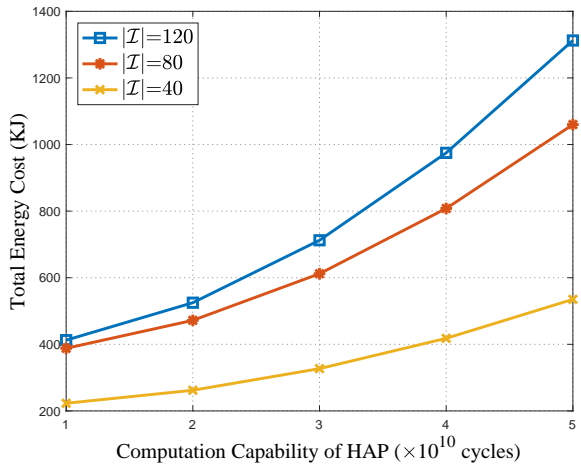}

}

\subfloat[Impact of UAV's computation capability ($C_{h}=5\times10^{10}\textrm{cycles}$).\label{sim:energycost-UAV}]{\centering

\includegraphics[scale=0.44]{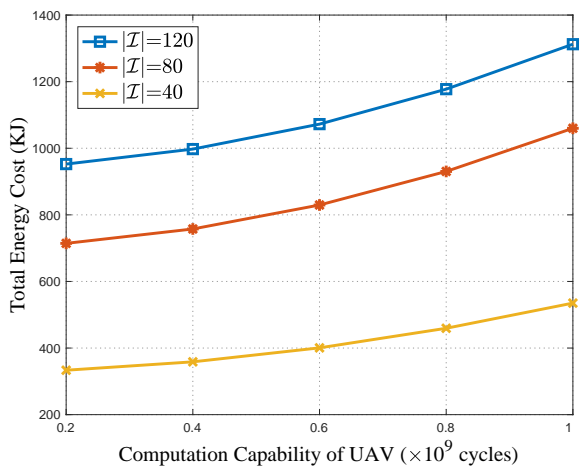}

}\caption{Impact of computation capability on energy cost.\label{sim:energycost}}
\end{figure}

Moreover, Fig. \ref{sim:energycost} reveals the effect from the computation
capability of HAPs and UAVs on total energy consumption. In particular,
from Fig. \ref{sim:energycost-HAP}, we can observe that the total
energy cost has an increment with the increasing of HAP's computation
capability, and similarly from Fig. \ref{eq:energycostUAV}, the total
energy consumption is increasing with the increment of UAV's computation
capacity. Such trends are in accordance with formula and (\ref{eq:energycostUAV})
and (\ref{eq:energycostHAP}). Besides, note that HAP's computation
capability variation has a stronger effect on the total energy cost
than UAVs, since the HAP equipped with large computing and energy
capacity provides service for more IoT devices.

\section{\textcolor{black}{Conclusions and Future Works}\label{sec:Conclusions}}

In this paper, we have investigated the hierarchical aerial computing
to serve the terrestrial IoT devices by cooperating HAPs and UAVs.
Two offloading schemes have be considered: IoT data being offloaded
to UAVs and computed at UAVs; IoT data being relayed by UAVs to HAPs
and computed at HAPs. The problem of maximizing total successful computed
data of IoT users has been formulated, which is in the form integer
programming and intractable to solve. Hence, we have presented the
computationally tractable matching game based algorithm to deal with
the data offloading from IoT to UAVs, and the external effect among
different IoT devices has also been tackled. Besides, a heuristic
algorithm regarding to the data offloading from UAVs to HAPs has been
designed, and after that, to take full advantage of the aerial resources,
an adjustment algorithm has been proposed. The complexity of the proposed
algorithms has been analyzed and numerical results have verified that
the proposed algorithms can efficiently achieve the near optimal solution,
compared with the exhaustive searching. Moreover, the advantages of
the IoT-UAV-HAP offloading scheme as well as the influence from various
network parameters have been analyzed, conducive to the resource management
in practical applications.\textcolor{black}{{} There exist a couple
of open issues to be addressed in the future works. First, the issue
of dynamic network with varying traffic load as well as the metric
of channel utilization will be considered in the aerial computing
networks. Second, we will further optimize the data rate and equipment
utilization in the aerial computing framework. Finally, we plan to
explore the mutual data offloading between UAVs through networking
to improve the calculation rate.}

\end{document}